# Harnessing Multifractality to Enhance Thermal Stability in Mixed-Phase Vanadium Oxide Thin Films

Abhijeet Das[1], Ram Pratap Yadav[2], Rashmi Roy Karmakar[3], Jyoti Jaiswal[3,*],

Sanjeev Kumar[3,*]

[1]Department of Bioengineering, Indian Institute of Science, Bangalore – 560012, India

[2]Department of Physics, Allahabad Degree College, University of Allahabad, Prayagraj – 211003, India

[3]Center for Advanced Research, Department of Physics, Rajiv Gandhi University, Rono Hills, Doimukh, Arunachal Pradesh - 791112, India

[*]**Corresponding Author:** jyoti.jaiswal@rgu.ac.in , sanjeev.kumar@rgu.ac.in

Vanadium oxide thin films exhibit temperature-driven electronic transitions desirable for sensing and microelectronic applications, yet their performance is often limited by thermal hysteresis. This study demonstrates that electronic stability is governed not simply by roughness or crystallinity but by a unique combination of surface morphological complexity and thermal hysteresis, revealed across films deposited with varying working pressure using Direct Current/Radio Frequency magnetron sputtering. Specifically, the film grown at 15 mTorr shows a distinct convergence of highest morphological complexity and lowest thermal hysteresis, exhibiting nearly reversible transport with activation energies ranging from 0.26 to 0.28 eV and negative temperature coefficients of resistance between −0.0337 and −0.035 $K^{-1}$. While conventional roughness metrics and mono-fractal parameters do not capture this behavior, multifractal detrended fluctuation analysis uncovers a pronounced peak in multifractality strength, which correlates inversely with thermal hysteresis. This highlights multifractality strength as a predictive descriptor of electronic stability, identifying a multiscale structural signature that enhances stress accommodation during thermal cycling. These results define an optimal deposition window and provide a morphology-guided pathway for developing thermally robust mixed-phase vanadium oxide films.

Vanadium Oxide ($V_xO_y$) thin films have garnered significant interest for their diverse applications in microelectronics, smart windows, temperature sensors, and memory devices owing to their metal-insulator transition (MIT) and thermochromic properties [1]. However, the practical implementation of these materials remains challenged by thermal hysteresis, which limits device reproducibility, reliability, and lifetime [2]. Understanding and minimizing this hysteresis is critical for advancing $V_xO_y$ thin film-based technologies.

$V_xO_y$ thin film exhibits a notable temperature coefficient of resistance ($TCR$), making it particularly valuable for applications in microbolometers, temperature sensors, and calorimeters [3, 4]. Recent studies have demonstrated that vanadium oxide films can achieve $TCR$ values from -2% to -3% $K^{-1}$, with corresponding resistivity values between 0.00819 and 0.04718 $\Omega \cdot cm$ [3]. These electrical properties can be precisely tuned through various synthesis techniques, including multilayer deposition methods that leverage the intermixing of vanadium and $V_xO_y$ layers. Nonetheless, while these properties are well documented, the relationship between film deposition or growth conditions, surface morphology, and electrical stability remains less well understood [1].

While numerous studies have investigated how deposition or growth parameters affect the crystallinity, stoichiometry, and electronic properties of $V_xO_y$ thin films, [5-8] the relationship between surface morphological complexity and electronic stability remains less understood. In addition, mixed-phase $V_xO_y$ thin films have been reported as candidates for optical applications [9].

Previous work has demonstrated that sputtering or working pressure significantly influences film microstructure and surface morphology through changes in adatom energy and mobility during deposition [10, 11]. However, conventional statistical parameters often provide insufficient characterization of the complex surface features (e.g., texture, roughness), facilitated by deposition parameters like working pressure, that emerge during thin film growth and their correlation with functional properties [10].

Fractal geometry-based analysis has emerged as a powerful tool for characterizing complex surfaces across multiple spatial and/or temporal scales and correlation with observable physical properties [10, 12]. More recently, multi-fractal analysis has extended this approach to capture the heterogeneity in scaling behavior, providing insights into local variations in surface complexity [10, 13]. Nevertheless, despite their potential, and to the best of our knowledge, these analytical methods have not been applied to understand the electronic stability of mixed-phase $V_xO_y$ semiconductor thin films deposited under varying working pressures.

In this letter, we report an investigation of mixed-phase $V_xO_y$ thin films deposited at varying working pressures with irregular intervals, i.e., 5, 15, 30, and 45 mTorr, using Direct Current/Radio Frequency (DC/RF) sputtering. For brevity, they are highlighted as S1, S2, S3, and S4, respectively, throughout the manuscript. We employ mono- and multi-fractal analysis to characterize surface morphological complexity and establish correlations with electronic properties, particularly thermal hysteresis, i.e., the difference in electronic properties between heating and cooling cycles. Our findings reveal that S2 exhibits a unique combination of multi-fractal characteristics that strongly correlate with thermal stability. Consequently, we propose that the multi-fractal spectrum width or multifractality strength could serve as a predictive metric for electronic stability, offering a unique approach to optimizing $V_xO_y$ thin films for device applications.

The detailed experimental steps, optimized parameters, and supporting characterizations are reported in our previous work [14]. Briefly, pristine $V_xO_y$ thin films were deposited on $1 \times 1$ cm$^2$ silicon substrates in a custom-built multilayer sputtering chamber. The substrates were ultrasonically cleaned using acetone, ethanol, and deionized water, then dried under nitrogen flow. During deposition, the chamber was evacuated to a base pressure of $4 \times 10^{-6}$ Torr and heated at 250 ºC. The target-to-substrate distance was maintained at 5.3 cm.

The resulting films consist of vertically aligned porous nanoflakes with a mixed phase composition, as confirmed by glancing-incidence X-ray diffraction (GIXRD) and X-ray photoelectron spectroscopy (XPS). The structure is dominated by $V_2O_5$, with inclusion of $VO_2$, both possessing an orthorhombic structure [14]. Surface morphology was characterized using field emission scanning electron microscopy (FESEM), and surface morphological features were analyzed from image analysis using conventional and fractal measures.

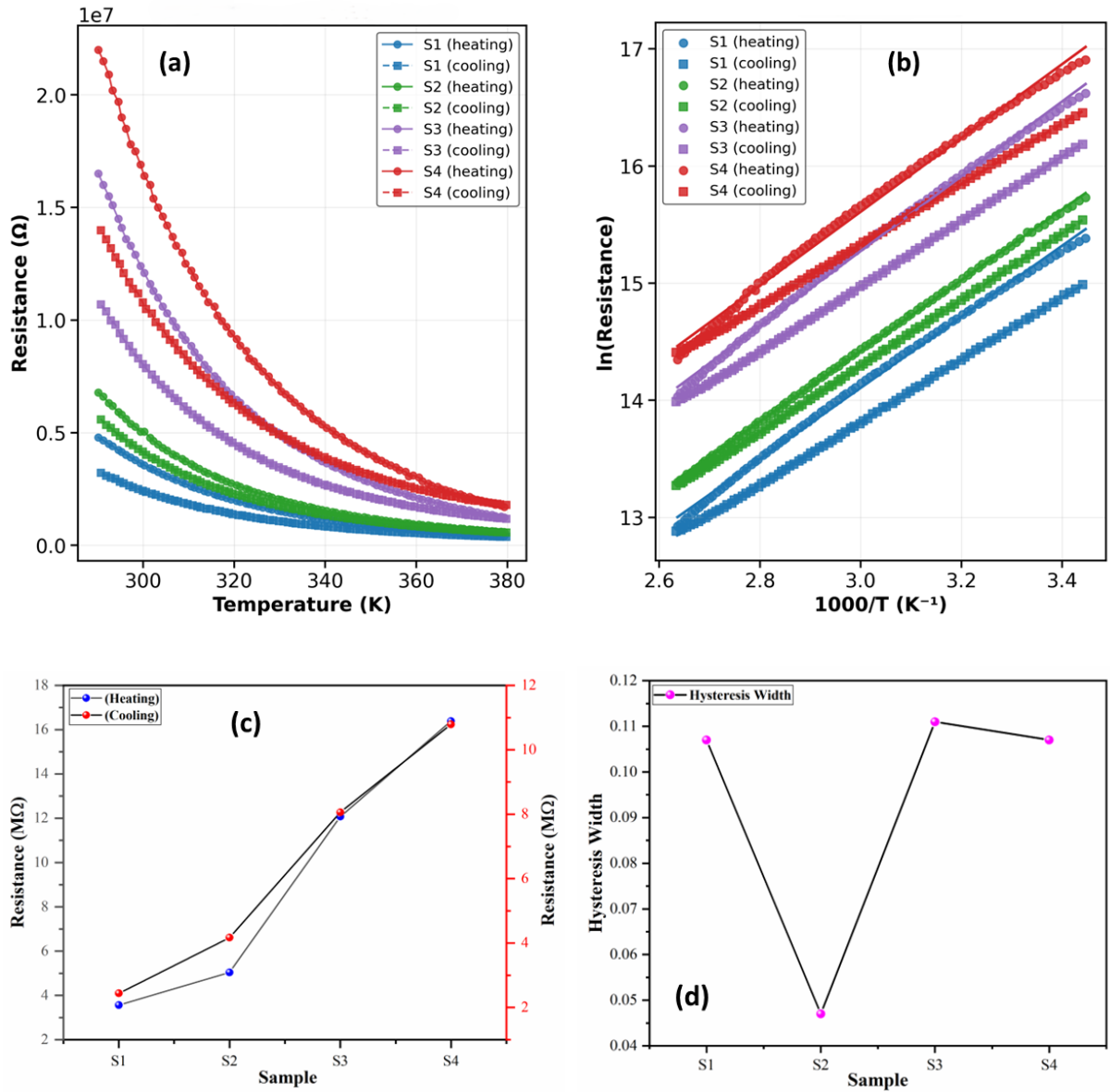

**Fig. 1 (a)** Resistance vs. temperature for S1 (5 mTorr), S2 (15 mTorr), S3 (30 mTorr), and S4 (45 mTorr) measured during a heating cycle (solid lines) and a subsequent cooling cycle (dashed lines), **(b)** Arrhenius plots- ln (Resistance) plotted versus 1000/Temperature (T) for heating (circle) and cooling (square). Linear fits were used to compute activation energies from the slope, **(c)** Room-temperature (300 K) resistance for each sample during heating and cooling cycles, **(d)** Thermal hysteresis expressed as the width between heating and cooling cycles at equivalent temperatures

Figure 1(a) shows the temperature-dependent variation in resistance during heating and cooling cycles for all the samples. Each film exhibits semiconducting behavior with decreasing resistance as temperature increases, consistent with thermally activated transport [15, 16]. The activation energies ($E_a$) extracted from Arrhenius plots (Fig. 1(b)) range from 0.2615±0.0013,

0.2601±0.0005, 0.2752±0.0013, and 0.2715±0.002 eV during the heating cycles and 0.229±0.0004, 0.2435±0.0002, 0.2386±0.0003, and 0.223±0.0003 eV during the cooling cycles for S1, S2, S3, and S4, respectively, with the coefficient of determination ($R^2$) being nearly 0.999. The $E_a$ was computed using Eq. (1-2):

$$\ln R = a + b\left(\frac{1000}{T}\right) \qquad (1)$$

$$E_a = bk_B \times 1000 \qquad (2)$$

Here, $R$ and $T$ denote resistance and temperature, $a$ and $b$ are fitting constants, $k_B = 8.617 \times 10^{-5}\ eV/K$ is the Boltzmann constant, and 1000 is the scaling constant.

Notably, the $E_a$ are significantly higher than the $E_a$ of 0.1 eV for pure-orthorhombic phase $V_2O_5$ nanoparticles [17]. The observed difference can be attributed to the investigated films' mixed-phase composition and additional scattering mechanisms at the $V_2O_5$ and $VO_2$ interfaces. In addition, the $E_a$ varied non-monotonically during the thermal cycles, likely due to the phase restructuring during thermal treatment.

A systematic increase in room temperature (300K) resistance with working pressure is observed, from 3.56 MΩ for S1 to 16.38 MΩ for S4 for the heating cycle, and from 2.44 MΩ (S1) to 10.78 MΩ (S4) for the cooling cycle (Fig. 1(c)). The trend is attributed to decreased bulk film density and increased overall porosity at higher sputtering pressure due to the lower adatom mobility [18]. Higher pressure deposition leads to more collisions in the plasma phase, reducing the energy of sputtered species reaching the substrate, which limits surface diffusion and promotes porous growth [18, 14]. Notably, the mechanism aligns with the established physics, where sputtering pressure significantly influences growth kinetics, film thickness, and porosity by altering the kinetic energy and particle volume within the deposition chamber.

The $TCR$ values at 300K are negative, with uncertainties negligible up to four decimals for all investigated thin films, i.e., -0.0337 to -0.035 $K^{-1}$ during heating and -0.0295 to -0.0288 $K^{-1}$ during cooling cycles, respectively, confirming semiconducting behavior. [3] The $TCR$ was computed using Eq. (3):

$$TCR = -\frac{E_a}{k_B T^2} \qquad (3)$$

Notably, $TCR$ magnitude exhibits minor variation with working pressure, showing a weak trend toward larger values in the heating cycle, indicating temperature dependence in film grown at higher pressures, likely due to increased grain boundary scattering and phase boundaries between $V_2O_5$ and $VO_2$. Contrastingly, it exhibits a minor decrement during the cooling cycle and can be correlated to thermally driven phase distribution, microstructural reorganization, or oxygen vacancy dynamics. [5, 14]

The average deviation in thermal hysteresis, quantified as the percentage difference in resistance between heating and cooling cycles across the common temperature range, shows a non-monotonic relationship with working pressure (Fig. 1(d)). The average deviation in thermal hysteresis ($\overline{H_d}$) was computed using Eq. (4-5):

$$H_d(T_i) = \frac{|R_{heating}(T_i) - R_{cooling}(T_i)|}{R_{heating}(T_i)} \times 100 \tag{4}$$

$$\overline{H_d} = \frac{1}{N}\sum_{i=1}^{N} H_d(T_i) \tag{5}$$

Here, $N$ is the total number of temperature bins, and $R_{heating}$ and $R_{cooling}$ represents the resistance during heating and cooling cycles, respectively, corresponding to the $i^{th}$ temperature ($T_i$).

While S1, S3, and S4 exhibited substantial deviation in hysteresis (25.45%, 26.08%, and 24.48%, respectively), S2 showed a markedly reduced value (12.36%) during heating and cooling cycles, respectively. The physical trend is reinforced by the integration of the $log_{10}(R) - T$ curves on the $1000/T$ scale using Trapezoidal and Simpsons 1/3 rule, yielding hysteresis widths of 0.107, 0.047, 0.111, and 0.107 from S1 to S4, respectively, where the notably smaller value for S2 reaffirms its minimal electronic hysteresis and superior thermal reversibility. This notable stability [19] is also reflected in the $E_a$ difference between heating and cooling cycles, where S2 exhibits the smallest difference (-6.38%) compared to S1 (-12.41%), S3 (-13.33%), and S4 (-17.84%), indicating that the cooling branch of S2 is closest to reversible behaviour. The anomalous behavior of S2 suggests an optimal deposition window in which the resulting film structure provides superior resistance to irreversible changes during thermal cycling. This finding can have implications for devices requiring thermal stability and reproducibility.

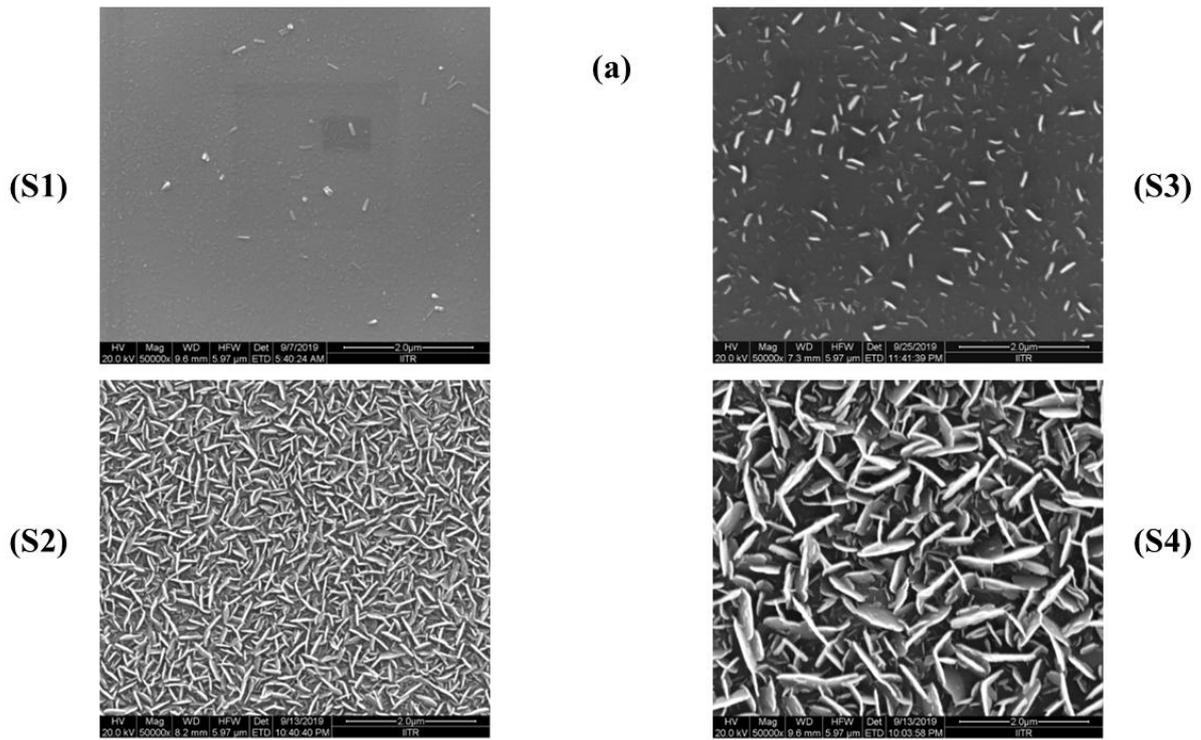

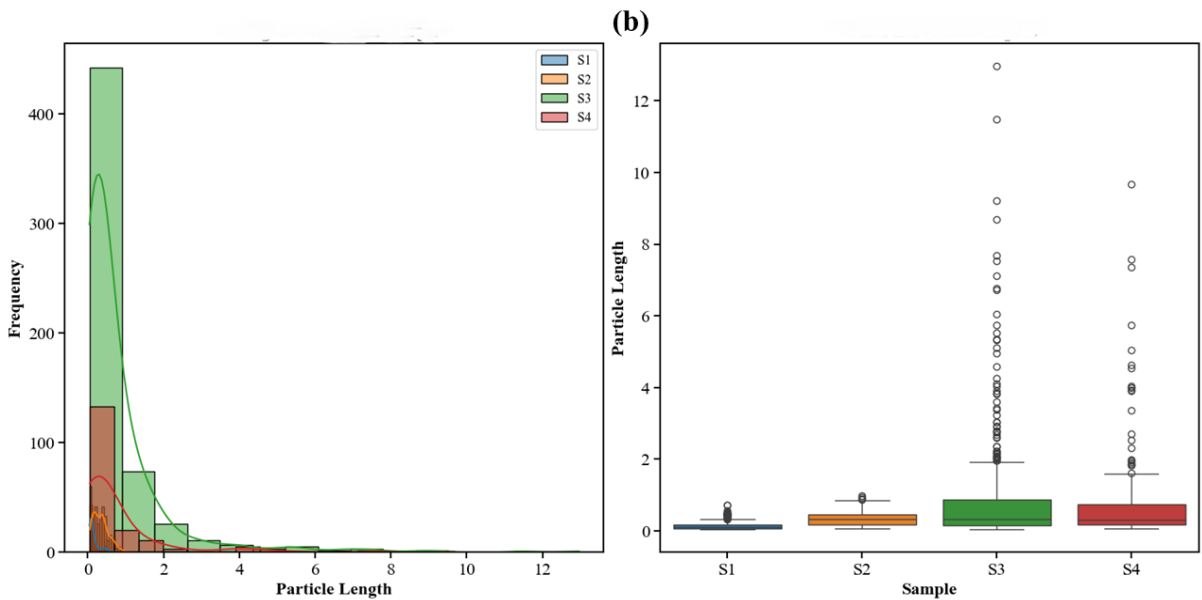

**Fig. 2 (a)** Representative FESEM micrographs of S1–S4 illustrating the morphological evolution with sputtering pressure, **(b)** Quantitative particle-length analysis derived from FESEM images - histogram of particle lengths for S1–S4; boxplots showing median, interquartile range, and outliers for each sample

The FESEM images (Fig. 2(a)) reveal distinct differences in surface morphology across the sequence of working pressures. Explicitly, S1 shows sparse, small nanostructures distributed across a relatively smooth surface, S2 exhibits an increased density of small, elongated nanostructures with relatively uniform distribution, and S3 demonstrates a dense network of interconnected nanoflakes covering almost the entire surface. In contrast, S4 shows larger, more defined nanoflake structures with increased spacing between features.

The observed morphological evolution with sputtering pressure can be explained as follows: at low sputtering pressure, the relatively high kinetic energy of sputtered atoms leads to limited nucleation sites but enables greater surface mobility of adatoms. This results in a film with higher bulk density but few structural features, as observed in S1. When pressure is increased to 15 mTorr (S2), a large number of aggregated nuclei combine due to the rapid nucleation process, which drives the formation of vertical grains with moderate surface feature density and uniformity, while maintaining reasonable bulk density. A further increase in pressure promotes the growth of oriented grains perpendicular to the substrate due to the anisotropic nature of vanadium oxide, leading to the highest surface feature density but decreased bulk density. This dual evolution, i.e., decreasing bulk film density with increasing working pressure, combined with a non-monotonic trend in surface feature density, creates a complex relationship between deposition conditions and observed film surface properties. Additionally, the increase in plasma density at higher sputtering pressure causes stronger bombardment of the substrate, which aids the development of adatoms into vertically aligned porous nanoflake structures. [14]

Quantitative analysis of particle lengths across samples (Fig. 2(b)) further elucidates these morphological differences. The histogram and box plots reveal that S1 has predominantly small particles with a narrow size distribution, while S2 shows slightly larger particles with a moderately broader distribution. S3 and S4 exhibit much broader and skewed size distributions with significant outliers, as evidenced by the numerous data points above the upper quartile in the box plots. Notably, S3 has the highest frequency of small-length particles (the tallest peak in the histogram), combined with a substantial number of large outliers.

To understand the underlying mechanisms of this enhanced stability, we characterized the surface morphological complexity using mono- and multi-fractal analysis. Although originally developed for time-series analysis, detrended fluctuation analysis (DFA) and multi-fractal detrended fluctuation analysis (MFDFA) can be extended to higher dimensions for analyzing spatial data such as surface morphology. For implementation, we treated the FESEM grayscale images as three-dimensional (3D) arrays, where pixel intensities represented the height information. The 2D-DFA and 2D-MFDFA were performed according to the methodology reported by Gu and Zhou [20].

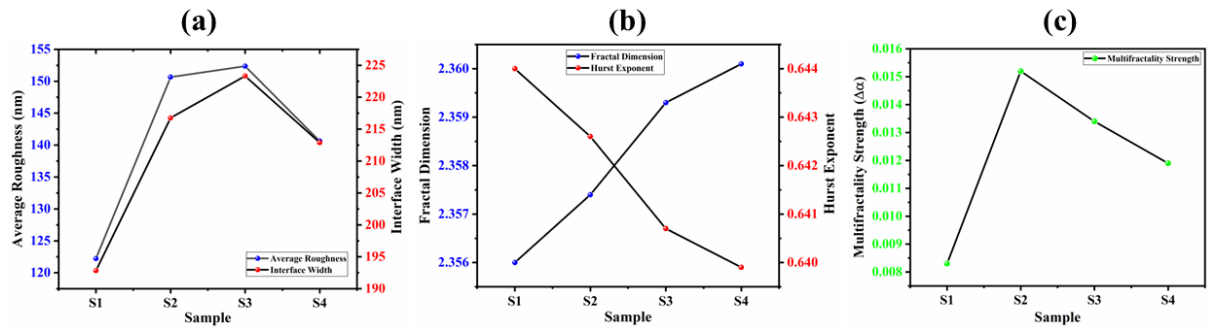

**Fig. 3** Variation in **(a)** conventional surface irregularity metrics, **(b)** mono-fractal parameters, and **(c)** multi-fractal parameter, obtained from image analysis, with working pressure

Conventional measurements of surface irregularity (Fig. 3(a)) show that average roughness increases from 122.22 nm for S1 to 152.36 nm for S3 before decreasing to 140.63 nm for S4. Interface width or root-mean-square roughness follows a similar non-monotonic trend. However, these parameters alone do not explain the unique stability of S2.

Mono-fractal analysis reveals a monotonic increase in fractal dimension from 2.356 (S1) to 2.3601 (S4), with a corresponding decrease in Hurst exponent from 0.644 to 0.6399 (Fig. 3(b)). This trend indicates progressively complex surface facilitated by enhancement in direction-dependent (or self-affine) space-filling growth patterns with increasing working pressure at the scale of measurement, but still fails to differentiate the unique behavior of S2.

Nevertheless, multi-fractal analysis provides a key insight. Fig. 3(c) shows the multifractality strength ($\Delta\alpha$) as a function of working pressure, revealing a noticeable maximum (= 0.0152) for the S2 compared to values of 0.0083, 0.0134, and 0.0119 for S1, S3, and S4, respectively. This parameter quantifies the heterogeneity of local scaling behaviors across different regions of the multifractal surface structure.

The strong negative correlation between multifractality strength and thermal average hysteresis, with the S2 sample positioned optimally at maximum multifractality strength and minimum hysteresis. Statistical analysis yields a Pearson correlation coefficient of -0.67, indicating a physically significant inverse relationship between these parameters.

This correlation reflects a connection between morphological complexity and electronic stability in mixed-phase $V_xO_y$ thin films. The enhanced multifractality strength for S2 indicates a surface with a substantial distribution of local scaling behavior of heights, suggesting structural heterogeneity across multiple length scales. This complex structure likely provides multiple pathways for adapting to stress during thermal cycles [21], thereby minimizing irreversible structural changes and reducing hysteresis.

The distinctive morphology of S2 in Fig. 2(a) and the particle size distribution in Fig. 2(b) support this interpretation. S2 exhibits a relatively uniform distribution of small elongated structures without the extreme variability in size as seen in S3 and S4, yet with greater feature density than S1. Thus, it is likely that this unique morphology, neither too sparse (like S1) nor too heterogeneous (like S3 and S4), appears optimal for electronic stability.

Unlike the pristine orthorhombic $V_2O_5$ phase reported by Dhoundiyal *et al* [17], the analyzed mixed-phase films exhibit more complex behavior due to the interaction between mixed $V_xO_y$ phases. However, the layered structure typical of pristine $V_2O_5$, confirmed by Raman spectroscopy [17], likely plays a significant role in the electronic transport and stability in the investigated thin films. Thus, it can be speculated that the presence of mixed phases introduces additional interfaces and boundary regions that can either enhance or degrade stability during thermal cycling, depending on their distribution and arrangement within the film structure.

At lower pressure (5 mTorr), higher energy deposition creates denser films but with less structural complexity to adapt to thermal stress, as evidenced from Fig. 2(a). The observation is consistent with the understanding that at low sputtering pressures, the reduced collision frequency in the gas phase results in sputtered atoms arriving at the substrate with higher kinetic energy, promoting densification but limiting the formation of complex structures. At higher pressures (30 and 45 mTorr), excessive porosity and defects dominate, compromising both conductance and stability, visible as the highly irregular and interconnected structures in Fig. 2(a). The increased scattering at these pressures reduces the kinetic energy of deposited species, limiting their surface mobility and resulting in more porous films with increased grain boundary scattering. This is corroborated by the observed increase in room temperature resistance and greater vulnerability to irreversible structural changes during thermal cycling. The 15 mTorr condition represents an optimal balance where sufficient energy enables adequate crystallization while maintaining beneficial morphological complexity. Here, we conjecture that moderate pressure creates a balance in the competing nucleation and surface diffusion processes, resulting in a film morphology with moderate bulk density and optimal surface feature characteristics that provide superior resistance to thermally induced structural changes.

The minimum and maximum values of the singularity strength parameter (α) further support the interpretation. The S2 sample exhibits the highest $α_{max}$ (2.0075) and lowest $α_{min}$ (1.9923) along with the highest $Δf$ (-1.763 × 10$^{-4}$), indicating the widest range of local scaling behaviors; along with the highest vertical complexity and more unsubstantial surface feature regions compared to denser regions [13]. The observed unique multi-fractal signature suggests that moderate sputtering pressure creates a self-organized complex surface in the morphological space that enhances electronic stability.

The reported findings can have technological implications: (i) the identification of 15 mTorr as a promising deposition working pressure for mixed-phase $V_xO_y$ thin films with enhanced thermal stability, and (ii) the observation of multi-fractal analysis as a useful indicative tool for electronic performance in functional oxide materials. Consequently, we conjectured that by quantifying the multifractality strength, it could be possible to predict hysteresis behavior without performing complete thermal cycling tests, facilitating rapid process optimization.

The correlation between multifractality and electronic stability may extend beyond $V_xO_y$ to other functional materials where thermal stability is crucial. Our work demonstrates that subtle aspects of surface morphology, captured by multi-fractal analysis but overlooked by conventional roughness measurements, can influence electronic properties relevant to device applications. Nevertheless, we acknowledge the limited sample size and emphasize the need for rigorous statistical analysis to substantiate the observation.

In summary, our integrated analysis of thermal hysteresis behavior and multi-fractal surface characteristics reveals a conceptual correlation between morphological complexity and electronic stability in mixed-phase $V_xO_y$ thin films. The identification of 15 mTorr as an optimal working pressure in DC/RF magnetron sputtering deposition for minimizing thermal hysteresis provides notable insights for fabricating $V_xO_y$-based devices with enhanced reliability and lifetime. Furthermore, the demonstrated utility of multi-fractal analysis as an indicative tool for morphology-facilitated electronic stability could facilitate new pathways for the rational design and optimization of functional oxide thin films for various electronic applications.

**Supplementary Material:** The raw resistance vs. temperature measurements for heating and cooling cycles, corresponding to all the samples, are included as a CSV file in the Supplementary Material.

**Declaration of Interest:** The authors declare that they have no known competing financial interests or personal relationships that could have appeared to influence the work reported in this paper.

**Data Availability:** The data that support the findings of this study are available within the article and its supplementary material.